\documentclass[prl,twocolumn,showpacs,superscriptaddress]{revtex4-1}

\usepackage{graphicx}
\usepackage{dcolumn,float}
\usepackage{amsmath,amssymb}
\usepackage{subfigure}
\usepackage{hyperref}
\usepackage{color}

\newcommand{\bra}[1]{\langle\left. #1 \right|}
\newcommand{\ket}[1]{\left| #1 \right.\rangle}
\newcommand{\braket}[1]{\langle #1 \rangle}

\newcommand{\rmi}{\mathrm{i}}
\newcommand{\rmd}{\mathrm{d}}
\newcommand{\rme}{\mathrm{e}}
\newcommand{\rbr}[1]{\left( #1 \right)}
\newcommand{\cbr}[1]{\left[ #1 \right]}
\newcommand{\tbr}[1]{\left\{ #1 \right\}}
\newcommand{\prodl}[2]{\prod\limits_{#1}^{#2}}
\newcommand{\sul}[2]{\sum\limits_{#1}^{#2}}
\newcommand{\pdiff}[2]{\frac{\partial #1}{\partial #2}}
\newcommand{\intg}[3]{\int\limits_{#1}^{#2}\rmd#3}
\newcommand{\sites}{L}
\newcommand{\V}[1]{{\bf #1}}
\newcommand{\gV}[1]{{\boldsymbol #1}}
\newcommand{\qfv}{\V{q}^{({\rm f})}}
\newcommand{\qiv}{\V{q}^{({\rm i})}}
\newcommand{\nfv}{\V{n}^{({\rm f})}}
\newcommand{\niv}{\V{n}^{({\rm i})}}
\newcommand{\nfj}[1]{n_{#1}^{({\rm f})}}
\newcommand{\nij}[1]{n_{#1}^{({\rm i})}}
\newcommand{\thetaij}[1]{\theta_{#1}^{({\rm i})}}
\newcommand{\thetaifj}[1]{\theta_{#1}^{({\rm i}/{\rm f})}}
\newcommand{\prefactor}{b}

\begin{document}

\title{Coherent Backscattering in Fock Space: \\ a Signature of Quantum Many-Body Interference in Interacting Bosonic Systems}

\newcommand{\RegensburgUniversity}{Institut f\"ur Theoretische Physik, 
Universit\"at Regensburg, D-93040 Regensburg, Germany}
\newcommand{\LiegeUniversity}{D\'epartement de Physique, 
Universit\'e de Li\`ege, 4000 Li\`ege, Belgium}

\author{Thomas Engl} \affiliation{\RegensburgUniversity}
\author{Julien Dujardin} \affiliation{\LiegeUniversity}
\author{Arturo Arg\"uelles} \affiliation{\LiegeUniversity}
\author{Peter Schlagheck} \affiliation{\LiegeUniversity}
\author{Klaus Richter} \affiliation{\RegensburgUniversity}
\author{Juan Diego Urbina} \affiliation{\RegensburgUniversity}

\begin{abstract}
We predict a generic manifestation of quantum interference in many-body bosonic systems 
resulting in a coherent enhancement of the average return probability in Fock space. 
This enhancement is both robust with respect to variations of external parameters and
genuinely quantum insofar as it cannot be described within mean-field approaches. 
As a direct manifestation of the superposition principle in Fock space,
it arises when many-body equilibration due to interactions sets in.
Using a semiclassical approach based on interfering paths in Fock space, we calculate the 
magnitude of the backscattering peak and its dependence on gauge fields that break time-reversal 
invariance. 
We confirm our predictions by comparing them to exact quantum evolution probabilities
in Bose-Hubbard models, and discuss the relevance of our findings in the context of 
many-body thermalization.
\end{abstract}

\pacs{03.65.Sq, 05.45.Mt, 67.85.-d, 72.15.Rn}

\maketitle

The existence of a superposition principle for quantum states is a cornerstone in our picture of 
the physical world, with observable implications in the form of coherent phenomena that have been 
experimentally demonstrated with impressive precision during the last century 
\cite{Rabi1,Rabi2,Rabi3}. 
Within the context of linear wave equations, quantum superposition effects 
represent particular cases of the general phenomenon of wave coherence.
For quantum systems described by the single-particle Schr\"odinger equation,
this analogy between quantum and classical waves was exploited to 
demonstrate coherent quantum effects, such as Anderson localization in disordered metals 
\cite{*[{For a recent review, see }] [{ reprinted in }] Ala, *Alb} or
coherent backscattering (CBS) \cite{Akk}, by using classical (in particular electromagnetic)
wave analogues \cite{Ph3,Ph2,Ph1}. 

In the quantum description of many-body systems, such an analogy between quantum dynamics
and classical wave phenomena does not hold. 
Within a first-quantized approach, the quantum mechanical description of a system 
of $N$ interacting particles in $D$ dimensions requires us to extend the space in 
which the Schr\"odinger field $\psi(\vec{r}_{1},\ldots,\vec{r}_{N},t)$ is defined 
to $ND$ dimensions.
We can still identify the quantum superposition principle with the linearity of the many-body 
Schr\"odinger equation, but the latter does no longer describe a classical wave 
in real $D$-dimensional space: Many-body quantum interference is a high-dimensional phenomenon.

\begin{figure}[t]
 \includegraphics[width=8.0cm,height=3.0cm]{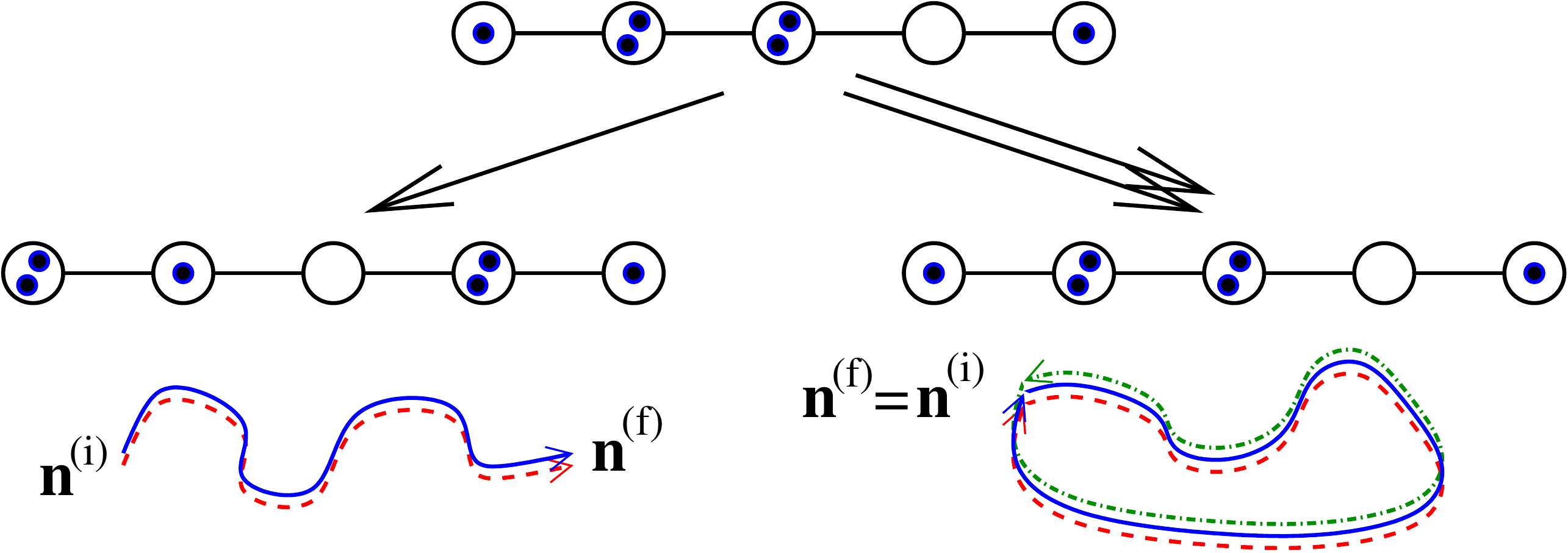}
\caption{(Color online) Illustration of Coherent Backscattering in Fock space. 
A many-body system, represented here by a Bose-Hubbard chain with $L=5$ sites and $N=6$ particles, 
is prepared in a well-defined initial Fock state 
$\mathbf{n^{({\rm i})}} = (n^{({\rm i})}_1, \ldots,n^{({\rm i})}_L)$ (upper part). 
After a given evolution time $t$, the final populations are found to be $\mathbf{n^{({\rm f})}}$
(lower parts). 
A solution $\gamma$ of the Gross-Pitaevskii equation joining $\mathbf{n^{({\rm i})}}$ with 
$\mathbf{n^{({\rm f})}}$ contributes with an amplitude 
$K_{\gamma} \simeq A_{\gamma}{\rm e}^{iR_{\gamma}}$ to this process, 
where $R_{\gamma}$ is a classical action. 
Under averaging, only pairs of identical Fock space trajectories yield a systematically 
nonvanishing contribution to the probability $P=|\sum_{\gamma}K_{\gamma}|^{2}$ 
when $\mathbf{n^{({\rm f})}} \ne \mathbf{n^{({\rm i})}}$ (left column). 
For $\mathbf{n^{({\rm f})}} = \mathbf{n^{({\rm i})}}$ (right column), however, constructive 
interference additionally arises if trajectories are paired with their time-reversed counterparts 
(dot-dashed green line). This gives rise to a coherent enhancement of the probability 
to detect the system in the initial Fock state after the evolution, as compared to other states
with comparable distributions of the population.} 
\label{fig:catch}
\end{figure}

This observation remains true even if we adopt a real-space description in terms of the 
{\it quantum} field $\hat{\psi}(\vec{r},t)$. 
Indeed, $\hat{\psi}(\vec{r},t)$ is an operator instead of a complex amplitude and 
does not represent a quantum state. 
Nevertheless, quantum fields are a suitable starting point to implement approximations 
to the full many-body problem in terms of classical wave equations for single particles, 
which effectively amounts to the substitution $\hat{\psi}(\vec{r},t) \to \psi(\vec{r},t)$ 
at the level of the Heisenberg equations of motion for $\hat{\psi}(\vec{r},t)$. 
This approach leads to the mean-field Gross-Pitaevskii equation (GPE)
\cite{MF1} and also to the Truncated Wigner method \cite{GarZol,SinLobCas01PRL,TW1,TW2} 
with its quantum (Wigner-Moyal) corrections \cite{Korsch1}. 
Since the GPE is a classical field equation, its nonlinearity does not pose a conflict with 
the linearity of quantum evolution. For the same reasons, however, the $\psi(\vec{r},t)$ field 
cannot represent a quantum state and its physical meaning requires further interpretation as a 
condensate fraction or order parameter. 
In particular, interference effects resulting from the (weakly nonlinear) GPE are 
{\it not} a consequence of many-body interference; they are {\it classical wave} 
effects proper of a classical field equation and are generically suppressed already for 
small interactions \cite{SF,*SFe,TW,HarO08PRL,Us}. 

In this paper we report a semiclassical description of the quantum mechanism responsible for 
many-body interference phenomena in interacting bosonic systems, which is schematically 
illustrated in Fig.~\ref{fig:catch}. Our approach is based on coherent sums over 
{\it multiple solutions} of the GPE in occupation number space. It predicts 
quantum coherence effects which are in quantitative agreement with
numerical simulations of Bose-Hubbard models describing cold atoms in optical lattices. 

Since many-body interference is most visible in genuine many-body observables (\textit{i.e.}, 
which cannot be written as sums of expectation values of single-particle operators),
we will study, as a representative example, the microscopic evolution probability
from one many-body state to another one.
Following standard techniques \cite{NO}, we introduce a discrete and orthogonal but otherwise 
arbitrary set of $L$ single-particle states ("orbitals") $\chi_1,\ldots,\chi_L$. 
The associated set of commuting bosonic occupation number operators $\hat{n}_{\alpha}$ 
has common (Fock) eigenstates $|{\bf n}\rangle=|n_{1},\ldots,n_{L}\rangle$ and integer eigenvalues 
$n_{\alpha}$ denoting the number of particles in each single-particle orbital $\chi_{\alpha}$. 
The transition probability between Fock states at time $t$ reads then
\begin{equation}
\label{eq:TrProb}
P({\bf n}^{({\rm f})},{\bf n}^{({\rm i})},t)=|\langle {\bf n}^{({\rm f})} |\hat{U}(t)| {\bf n}^{({\rm i})} \rangle|^{2},
\end{equation}
with $\hat{U}(t)\equiv\exp(-i\hat{H}t/\hbar)$ and the general many-body Hamiltonian
exhibiting two-body interaction
\begin{equation}
\hat{H}=\sum_{\alpha\beta}h_{\alpha\beta}\hat{a}_{\alpha}^{\dagger}\hat{a}_{\beta}+\frac{1}{2}\sum_{\alpha\beta\eta\sigma}V_{\alpha\beta\eta\sigma}\hat{a}_{\alpha}^{\dagger}\hat{a}_{\beta}  \hat{a}_{\eta}^{\dagger}\hat{a}_{\sigma},
\end{equation}
which is expressed in terms of the ladder operators $\hat{a}_{\alpha} (\hat{a}_{\alpha}^{\dagger})$ 
that annihilate (create) a particle in the orbital $\chi_\alpha$ 
(with $\hat{n}_{\alpha}=\hat{a}_{\alpha}^{\dagger}\hat{a}_{\alpha}$).
We shall later on refer to the more specific case of a Bose-Hubbard (BH) model describing, 
e.g., cold atoms in optical lattices, which corresponds to the choice  
\begin{eqnarray}
\label{eq:BH}
\hat{H}_{\rm BH}&=&\sum_{\alpha}\epsilon_{\alpha}\hat{n}_{\alpha}- J\sum_{\alpha}({\rm e}^{i\phi}\hat{a}_{\alpha}^{\dagger}\hat{a}_{\alpha+1}+{\rm e}^{-i\phi}\hat{a}_{\alpha+1}^{\dagger}\hat{a}_{\alpha}) \nonumber \\ &+&\frac{U}{2}\sum_{\alpha}\hat{n}_{\alpha}(\hat{n}_{\alpha}-1).
\end{eqnarray}

Our calculation (see \cite{*[{See }] [{}] Tom} for details) is based on an asymptotic 
expansion of the many-body propagator,
\begin{equation}
\label{eq:MaElU}
K^{\rm sc}({\bf n}^{({\rm f})},{\bf n}^{({\rm i})},t) \simeq 
\langle {\bf n}^{({\rm f})} |\hat{U}(t)| {\bf n}^{({\rm i})} \rangle \;, 
\end{equation}
which is formally valid for $n_{\alpha}^{({\rm i,f})} \gg 1$.
To this end, we have to consider all solutions (indexed by $\gamma$)
\begin{equation}
\label{eq:solutions}
{\boldsymbol \psi}^{(\gamma)}(s)\equiv
{\boldsymbol\psi}^{(\gamma)}(s;{\bf n}^{({\rm f})},{\bf n}^{({\rm i})},t)
\equiv[\psi^{(\gamma)}_{1}(s), \ldots,\psi^{(\gamma)}_{L}(s)]
\end{equation}
of the mean-field Gross-Pitaevskii equation (GPE)
\begin{equation}
\label{eq:GP}
i\hbar\frac{\partial}{\partial s} \psi_{\alpha}=\sum_{\beta}h_{\alpha\beta}\psi_{\beta}+\sum_{\beta\eta\sigma}V_{\alpha\beta\eta\sigma}\psi_{\beta}\psi_{\eta}^{*}\psi_{\sigma}
\end{equation}
that instead of initial conditions satisfy the bilateral boundary (or shooting) conditions
$|\psi_{\alpha}(0)|^2=n_{\alpha}^{({\rm i})}+1/2$ and
$|\psi_{\alpha}(t)|^2=n_{\alpha}^{({\rm f})}+1/2$
and have $\arg{\psi_{\alpha=1}(0)}=0$ \cite{Tom}.
In terms of those solutions ${\boldsymbol \psi}^{(\gamma)}(s)$, 
the semiclassical propagator is then expressed as
\begin{equation}
\label{eq:vVG}
K^{\rm sc}({\bf n}^{({\rm f})},{\bf n}^{({\rm i})},t)=
\sum_{\gamma}A^{(\gamma)}\exp[i R^{(\gamma)}+i\pi \Phi^{(\gamma)} / 4]
\end{equation}
where, for each solution, the semiclassical amplitude
\begin{equation}
\label{eq:As}
A^{(\gamma)}({\bf n}^{({\rm f})},{\bf n}^{({\rm i})},t)=\sqrt{\left|{\rm det'}\frac{1}{2 \pi }\frac{\partial^{2} R^{(\gamma)}({\bf n}^{({\rm f})},{\bf n}^{({\rm i})},t)}{\partial{\bf n}^{({\rm f})}\partial {\bf n}^{({\rm i})}}\right|},
\end{equation}
is given by the (dimensionless) classical action 
\begin{eqnarray}
\label{eq:Rs}
&&R^{(\gamma)}({\bf n}^{({\rm f})},{\bf n}^{({\rm i})},t) =\\ &&\int_{0}^{t}\left(\sum_{\alpha}\theta_{\alpha}^{(\gamma)}(s) \dot{I}_{\alpha}^{(\gamma)}(s)-H[{\boldsymbol \psi}^{(\gamma)}(s)]/\hbar \right)ds \nonumber
\end{eqnarray}
with $\psi_{\alpha}^{(\gamma)}(s)\equiv\sqrt{I_{\alpha}^{(\gamma)}(s)}\exp[i \theta_{\alpha}^{(\gamma)}(s)]$ and
\begin{equation}
H({\boldsymbol \psi})=\sum_{\alpha\beta}h_{\alpha\beta}\psi_{\alpha}^{*}\psi_{\beta}+\frac{1}{2}\sum_{\alpha\beta\eta\sigma}V_{\alpha\beta\eta\sigma}\psi_{\alpha}^{*}\psi_{\eta}^{*} \psi_{\beta}\psi_{\sigma}
\end{equation}
the classical (mean-field) Hamiltonian. The Morse index $\Phi^{(\gamma)}$ counts the number of
conjugate points along the trajectory $\gamma$.
As indicated by ${\rm det'}$, the derivatives in Eq.~(\ref{eq:As}) are to be taken 
with respect to $n_2^{({\rm i/f})},\ldots,n_L^{({\rm i/f})}$ with 
$n_1^{({\rm i/f})}$ being fixed
by the total number of particles \cite{Tom}.

The heuristic use of $H({\boldsymbol \psi})$ as the classical limit in bosonic systems has a 
long history \cite{Hcl} and lies behind most studies of the quantum-classical correspondence 
in Bose-Hubbard models \cite{TW1, TW2,Korsch1,Sc}. 
However, a rigorous approach in which a semiclassical propagator is constructed by 
a stationary phase analysis of the {\it exact} path-integral representation of $\hat{U}(t)$ 
in the spirit of the van~Vleck-Gutzwiller approach for first-quantized systems 
\cite{*[{See }] [{ and references therein}] vVG} was missing in previous studies.
Importantly, our propagator $K^{\rm sc}$ is valid also if the classical limit is non-integrable, 
thus going beyond the successful WKB method of Refs.~\cite{2sites1,2sites2} for $L=2$ 
and the EBK approach of Ref.~\cite{Sc} for $L=3$. 
Contrary to previous classical and quasiclassical approaches (including
the standard implementations of the Truncated Wigner method 
\cite{GarZol,SinLobCas01PRL,TW1,TW2}), the classical information appears in Eq.~(\ref{eq:vVG})
in terms of a boundary value problem generally exhibiting many solutions, 
instead of an initial value problem with a unique solution. 

Substituting Eqs.~(\ref{eq:MaElU}) and (\ref{eq:vVG}) into Eq.~(\ref{eq:TrProb}) yields
\begin{equation}
\label{eq:TrProbSc}
P({\bf n}^{({\rm f})},{\bf n}^{({\rm i})},t)=
\sum_{\gamma \gamma'}A^{(\gamma)} A^{(\gamma')}{\rm e~}^{i (R^{(\gamma)}-R^{(\gamma')})}.
\end{equation}
From the typical scaling $R^{(\gamma)}-R^{(\gamma')} \propto N$ of the action differences,
the contributions to the double sum in Eq.~(\ref{eq:TrProbSc}) contain a
large number of highly oscillatory terms that tend to cancel each other. 
Averaging, e.g., over a disorder potential that is contained in the matrix elements $h_{\alpha\beta}$
then selects contributions from those pairs of classical solutions 
that generically exhibit action quasi-degeneracies:
$R^{(\gamma)}-R^{(\gamma')} \sim 0$. 
The first non-vanishing contribution to the average transition probability 
(which is denoted by a horizontal bar, as any other averaged expression) is then given 
by the incoherent ($\gamma=\gamma'$) part of the double sum, 
\begin{eqnarray}
\label{eq:TrProbDg}
&&\bar{P}^{\rm cl}({\bf n}^{({\rm f})},{\bf n}^{({\rm i})},t)=
\sum_{\gamma}\overline{|A^{(\gamma)}|^{2}} \\
&&=\int_{0}^{2\pi}\frac{d\theta_2}{2 \pi} \ldots \int_{0}^{2\pi}\frac{d\theta_L}{2 \pi}
\overline{\prod_{\alpha=2}^L
\delta[n_{\alpha}^{({\rm f})}-|\psi_{\alpha}({\bf n}^{({\rm i})},{\boldsymbol \theta},t)|^{2}]}
\nonumber  
\end{eqnarray}
where ${\boldsymbol \psi}({\bf n}^{({\rm i})},{\boldsymbol \theta},t)$ is the unique solution of the 
GPE (\ref{eq:GP}) with initial conditions satisfying 
$|\psi_{\alpha}^{({\rm i})}|^{2}=n_{\alpha}^{({\rm i})}+1/2$ 
and $\arg \psi_{\alpha}^{({\rm i})}=\theta_{\alpha}$ with $\theta_{\alpha=1}=0$ 
\footnote{In order to derive the second line of \unexpanded{Eq.~(\ref{eq:TrProbDg})},
we use the classical identity 
\unexpanded{${\boldsymbol \theta}^{({\rm i/f})}=
\mp\partial R({\bf n}^{({\rm f})},{\bf n}^{({\rm i})},t)/\partial {\bf n}^{({\rm i/f})}$} 
(minus for ${\boldsymbol \theta}^{({\rm i})}$ and plus for 
${\boldsymbol \theta}^{({\rm f})}$) for each trajectory \unexpanded{$\gamma$}.}. 
Eq.~(\ref{eq:TrProbDg}) is the averaged transition probability 
obtained using the classical Truncated Wigner method 
\footnote{In the Truncated Wigner approach, one propagates, instead of a single trajectory, 
full initial manifolds in phase space $({\bf n},{\bf \theta})$ 
representing the initial quantum state. 
For the case of the transition probability, the manifold 
\unexpanded{$\delta(\bf{n}-\bf{n}^{({\rm i})})$} is classically propagated 
and projected at time \unexpanded{$t$} over \unexpanded{$\delta(\bf{n}-\bf{n}^{({\rm f})})$}, 
thus giving \unexpanded{$\bar{P}^{\rm cl}$}.}. 

Having identified $\bar{P}^{\rm cl}$ as the classical probability, any other 
robust contribution to $\bar{P}$ is necessarily a signature of many-body quantum interference. 
As shown schematically in Fig.~\ref{fig:catch}, having exact and generic action degeneracies
for $\gamma \ne \gamma'$ requires the presence of \emph{time-reversal invariance} (TRI), 
which means that for each solution ${\boldsymbol \psi}^{(\gamma)}(s)$ of the GPE 
one can find suitable phases $\omega_{\alpha}$ such that its time-reversal partner 
${\boldsymbol \psi}^{({\cal T}\gamma)}(s)$, with 
\begin{equation}
\label{eq:TRI}
\psi_{\alpha}^{({\cal T}\gamma)}(s;{\bf n}^{({\rm f})},{\bf n}^{({\rm i})},t)\equiv
{\rm e~}^{i\omega_{\alpha}}
[\psi_{\alpha}^{(\gamma)}(t-s;{\bf n}^{({\rm i})},{\bf n}^{({\rm f})},t)]^{*},
\end{equation}
is also a solution of the GPE but with the initial and final conditions interchanged. 
In that case, it follows from Eq.~(\ref{eq:Rs}) that $\gamma$ and ${\cal T}\gamma$ 
have the same classical actions and semiclassical amplitudes. 
Obviously, as the trajectories $\gamma, \gamma'$ in the double sum
(\ref{eq:TrProbSc}) 
refer to a specific ${\bf n}^{({\rm i})}$ and a specific ${\bf n}^{({\rm f})}$, 
a pairing $\gamma'={\cal T}\gamma$ is only possible if 
${\bf n}^{({\rm f})}={\bf n}^{({\rm i})}$ 
\footnote{Loop corrections \cite{KDa,*KDb} can be shown to identically vanish in this case.}.
Finally, as long as generically $\gamma \ne {\cal T}\gamma$, we obtain as our main result
\begin{equation}
\label{eq:TrProbDgTr}
\bar{P}({\bf n}^{({\rm f})},{\bf n}^{({\rm i})},t) \simeq 
\left(1+\delta_{{\bf n}^{({\rm f})},{\bf n}^{({\rm i})}} \delta_{\rm TRI}\right)
\bar{P}^{\rm cl}({\bf n}^{({\rm f})},{\bf n}^{({\rm i})},t)
\end{equation}
with $\delta_{\rm TRI} \equiv 1$ in the presence of TRI and $0$ otherwise.
This reflects \emph{coherent backscattering (CBS) in Fock space}, \textit{i.e.},
a coherent enhancement of the averaged quantum probability of return in Fock space 
over the classical value due to quantum many-body interference. 
Resulting from phase cancellations among oscillatory functions, 
this enhancement is a non-perturbative effect in the effective Planck constant 
$\hbar_{\rm eff} \sim N^{-1}$.

\begin{figure}[ttt]
 \includegraphics[width=8cm,height=5.5cm]{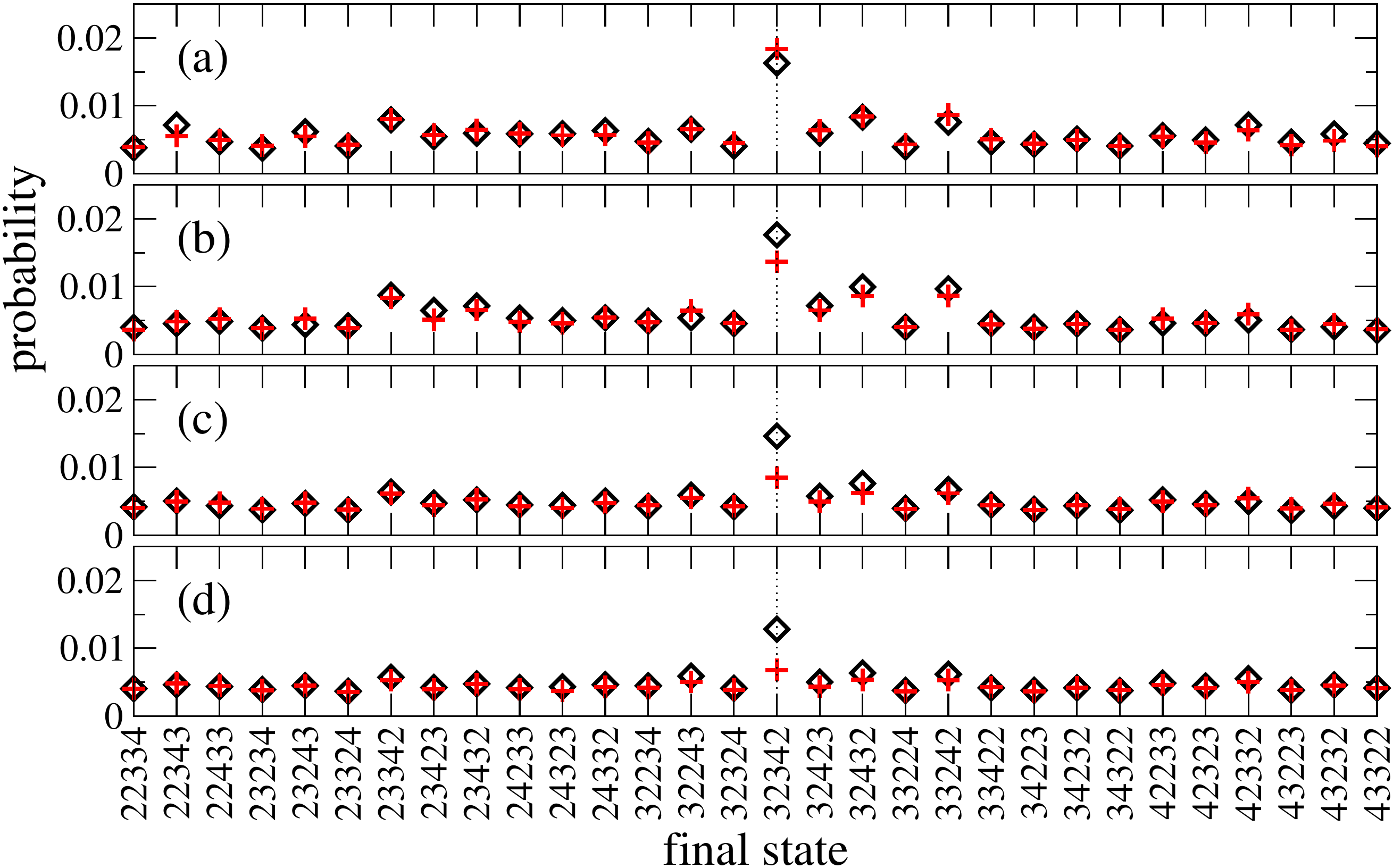}
\caption{(Color online) Average quantum (black diamonds) and classical (red crosses) 
evolution probabilities in Fock space for a Bose-Hubbard chain ($L=5,N=14$), 
at different evolution times (a) $t = 1.5\tau$, (b) $t = 2.5\tau$, (c) $t = 5\tau$, 
(d) $t = 10\tau$, with $\tau\equiv\hbar/J$. 
The average was performed over an ensemble of $10^{3}$ realizations of on-site energies 
$\epsilon_\alpha \in [0,10J]$, with interaction strength $U=4J$. 
The probabilities are displayed for the set of Fock states $\mathbf{n}^{(\rm f)}$ 
having the same total interaction energy as $\mathbf{n}^{({\rm i})} = (3,2,3,2,4)$ 
(marked by the vertical dashed line). 
Although for $t \gtrsim 5\tau$ equilibration in Fock space generally sets in, 
the quantum backscattering probability to $\mathbf{n}^{({\rm f})} = \mathbf{n}^{({\rm i})}$ 
is, in accordance with Eq.~(\ref{eq:TrProbDgTr}), systematically enhanced by about a factor two 
as compared to other final states $\mathbf{n}^{({\rm f})} \neq \mathbf{n}^{({\rm i})}$ 
and to its classical prediction. 
\label{fig:NumResI}}
\end{figure}

To confirm our result, Eq.~(\ref{eq:TrProbDgTr}), we performed extensive numerical calculations 
for the BH model defined in Eq.~(\ref{eq:BH}) for chain and ring topologies. 
We defined our ensemble average through independent variations of the on-site energies
$\epsilon_{\alpha}$, which are randomly selected from the interval $0 < \epsilon_{\alpha} < W$.
Taking advantage of the literature concerned with classical equilibration and chaos for 
this kind of Hamiltonians \cite{CBH1,CBH2,CBH3}, we fixed the numerical values of the free 
parameters $U/J$ and $W/J$ such that the classical phase space has a dominant chaotic component. 
The time $t$ is measured in units of the inverse Rabi 
frequency, $\tau\equiv\hbar/J$, between neighbouring sites.
The quantum transition probability $\bar{P}({\bf n}^{({\rm f})},{\bf n}^{({\rm i})},t)$ 
is then computed with a Runge-Kutta solver, using the exact quantum propagation of the initial 
state in full Fock space, followed by the disorder average over the on-site energies. 
The classical probability $\bar{P}^{\rm cl}({\bf n}^{({\rm f})},{\bf n}^{({\rm i})},t)$, 
on the other hand, is directly computed from Eq.~(\ref{eq:TrProbDg}) where, for a given 
random choice of the on-site energies, 
${\boldsymbol \psi}({\bf n}^{({\rm i})},{\boldsymbol\theta},t)$ 
is determined by the numerical solution of the $L$-dimensional GPE. 

In Fig.~\ref{fig:NumResI} we show the time dependence of $\bar{P}$ and $\bar{P}^{\rm cl}$ 
as a function of ${\bf n}^{({\rm f})}$ for the BH model (\ref{eq:BH}) with $L=5$, $N=14$ and a
chain topology (\textit{i.e.}\ the site $1$ is not connected to the site $L$ by a single 
hopping matrix element) starting from a generically chosen initial state 
${\bf n}^{({\rm i})} = (3,2,3,2,4)$.
After a transient time regime in which quantum and classical results resemble each other, 
the quantum transition probabilities clearly display, for $t \gtrsim 5 \tau$, a CBS peak at the 
initial state ${\bf n}^{({\rm i})}$ on top of a roughly constant background, 
in quantitative agreement with Eq.~(\ref{eq:TrProbDgTr}).
This peak is not reproduced by the classical probabilities ruling out short-time effects 
or self-trapping due to rare realizations of the random on-site energies 
as possible alternative origins of the enhancement. 

\begin{figure}[ttt]
 \includegraphics[width=8cm,height=6.5cm,angle=0]{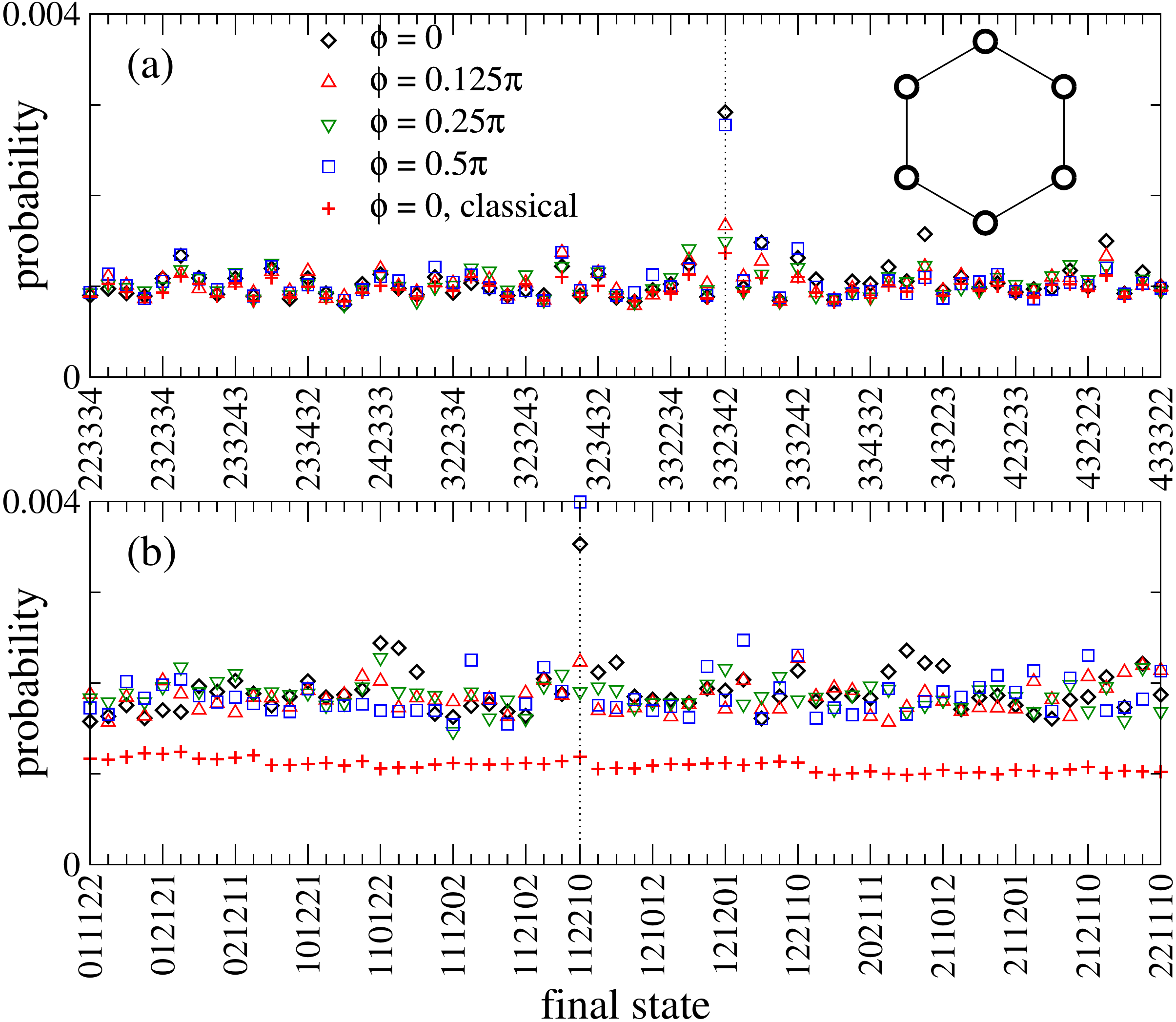}
\caption{(Color online) 
Evolution probability $\bar{P}({\bf n}^{({\rm f})},{\bf n}^{({\rm i})},t)$ 
in Fock space averaged over random on-site energies $\epsilon_{\alpha} \in [0,W]$ 
for a Bose-Hubbard ring of $L=6$ sites (see inset). 
We show results for evolution times $t = 10 \tau$, hopping phases $\phi=0,\pi/8,\pi/4,\pi/2$ 
(black diamonds, red upper triangles, green lower triangles, blue squares) and initial states 
${\bf n}^{({\rm i})}$ indicated by a vertical line. 
In (a) we have $N=17$ particles, with interaction strength $U=4J$ and $W=10J$. 
The lower (more ``quantum'') panel (b) has $N=7$, $U=J$ and $W=2J$. 
In both cases, the breaking of time-reversal invariance for $\phi=\pi/8,\pi/4$ 
destroys the coherent enhancement of the backscattering probability to the initial state. 
In the semiclassical regime (a) the evolution probabilities globally agree with the 
classical prediction for $\phi=0$ (red crosses), 
while they significantly exceed the latter in the quantum regime (b).}
\label{fig:NumResII}
\end{figure}

Fig.~\ref{fig:NumResII} shows the CBS peak at $t=10\tau$ for a BH ring (see inset)
with $L=6$ sites (in which a hopping matrix element connects the site $1$ to the site $L$) 
in the presence of nonvanishing hopping phases $\phi$ [see Eq.~(\ref{eq:BH})
which break TRI.
We clearly see the suppression of the CBS peak in the absence of TRI at $\phi=\pi/8$ and $\pi/4$.
For $\phi=\pi/2$, on the other hand, TRI is again established using $\omega_{\alpha}=\alpha \pi$
in Eq.~(\ref{eq:TRI}) and the CBS peak re-appears.

From the experimental point of view, CBS in many-body space could possibly
be observed with ultracold bosonic atoms. 
The specific ring geometry of Fig.~\ref{fig:NumResII} could be realized
in hexagonal (graphene) optical lattices \cite{PT1}. 
By means of a red-detuned laser beam which is tightly focused perpendicular to the graphene lattice, 
an individual hexagon could be isolated from the lattice. 
Displacing the focus of the laser beam with respect to the geometric center of this hexagon 
would allow one to load this ring in a non-uniform manner, \textit{i.e.}\ 
such that the atomic populations differ from site to site. 
While the ring is initially to be loaded in the deep Mott insulator regime, 
in which inter-site hopping along the ring is negligibly small, 
a sudden increase of the hopping strength at time $t=0$ will make the 
atoms propagate along the ring. 
At a given final propagation time, the system would have to be quenched back 
to the Mott regime and the atomic populations on the individual sites would 
have to be measured using, e.g., high-resolution imaging techniques \cite{PT3a,*PT3b}. 
In order to obtain the final probability distribution in Fock space with good statistical accuracy, 
optical disorder \cite{PT4} can be used to randomly vary the on-site energies in a controlled 
manner, and an artificial gauge field \cite{PT5} could be induced in order to break TRI
The lower panel of Fig.~\ref{fig:NumResII} displays the numerically computed Fock state 
probabilities on such a ring for the initial state ${\bf n}^{({\rm i})} = (1,1,2,2,1,0)$
\footnote{In order to prevent three-body losses, it is preferrable to avoid populations 
of more than two atoms per site.}. 
It clearly displays the CBS enhancement despite the fact that this initial state is far from 
semiclassical ($N/L \simeq 1$).

Our results represent a further step in the active field of thermalization in closed 
many-body systems \cite{CBH1,Ther1,Ther2,Ther3,Ther4}.
Indeed, Eq.~(\ref{eq:TrProbDgTr}) shows that in equilibrium, even in the semiclassical limit 
and when the classical system displays full ergodicity, many-body quantum interference generically 
inhibits {\it quantum} ergodicity in equilibrium, \textit{i.e.}
\begin{equation}
\bar{P}({\bf n}^{({\rm f})},{\bf n}^{({\rm i})},t\gg t_{{\rm eq}})\ne 1/{\cal N}_{\rm acc}
\end{equation}
where ${\cal N}_{\rm acc}\equiv {\cal N}_{\rm acc}({\bf n}^{({\rm i})})$ 
is the number of final Fock states that are energetically accessible to $ {\bf n}^{({\rm i})}$.
This result, however, is not in conflict with signatures of many-body thermalization at the level of 
single-particle observables, like the equilibration towards uniform occupation numbers reported in 
Ref.~\cite{Ther4}. 
As a matter of fact, Eq.~(\ref{eq:TrProbDgTr}) has an 
extremely small effect on such single-particle observables, as can be easily seen by calculating 
the averaged values $\bar{n}_{\alpha}(t)$, which gives the expected uniform behaviour
\begin{equation}
\bar{n}_{\alpha}(t, {\bf n}^{({\rm i})})=(N/L)+{\cal O}(1/{\cal N}_{\rm acc})
\end{equation}
in the regime of classical ergodicity.
However, for a genuine many-body observable like the inverse participation ratio 
${\rm IPR}(t)\propto\sum_{{\bf n}}P({\bf n},{\bf n},t)$, one may obtain 
a strong quantum correction, 
$\overline{{\rm IPR}}(t)=2\ \overline{{\rm IPR}}^{\rm cl}(t)$, in the presence of TRI.

To summarize, we presented a semiclassical approach in the van Vleck-Gutzwiller 
spirit, using sums over interfering paths that solve a classical mean field equation, 
which successfully captures genuine quantum interference in interacting bosonic systems. 
We used this approach to predict a clear-cut quantum (and genuinely many-body) 
effect, namely the coherent enhancement of the return probability in Fock space. 
Our predictions are fully confirmed by extensive simulations of Bose-Hubbard models with different 
topologies, even in the deep quantum regime where experimental observation 
using ultracold atoms is possible.

{\bf Acknowledgements}: We thank A.~Altland, T.~Guhr, B.~Gutkin, F.~Haake, 
M.~Oberthaler, W.~Strunz, and T.~Wellens for valuable discussions. 
This work was financially supported by the Deutsche Forschungsgemeinschaft within 
the DFG Research Unit FOR760 as well as by a ULg research grant for T.E.\ at the 
Universit\'e de Li\`ege.

\newpage

\onecolumngrid

\begin{center}
\textbf{\large Supplementary material for \\[0.2em]
``Coherent Backscattering in Fock Space: \\ 
a Signature of Quantum Many-Body Interference in Interacting Bosonic Systems''} \\[1em]
Thomas Engl,$^1$ Julien Dujardin,$^2$ Arturo Arg\"uelles,$^2$
Peter Schlagheck,$^2$ Klaus Richter,$^1$ and Juan Diego Urbina$^1$ \\[0.5em]
\textit{$^1$\RegensburgUniversity \\ $^2$\LiegeUniversity \\[0.5em]}
\end{center}

\twocolumngrid

The semiclassical propagator
cannot be directly obtained in Fock state representation, 
since the Fock states form a discrete basis rather than a continuous 
one as required by the path integral formalism. 
To solve this problem, we first derive a 
semiclassical propagator in quadrature representation and then
project the result on Fock states.

The quadrature eigenstates $\ket{\V{q}} \equiv \ket{q_1,\ldots,q_L}$ and
$\ket{\V{p}} \equiv \ket{p_1,\ldots,p_L}$ are defined as the eigenstates of 
linear hermitian combinations of the creation and annihilation operators
associated with the single-particle orbitals 
$\chi_\alpha$ ($\alpha=1,\ldots,L$), \textit{i.e.,}
\begin{align}
 \prefactor\rbr{\hat{a}_{\alpha}+\hat{a}_{\alpha}^\dagger}\ket{\V{q}}&=q_{\alpha}\ket{\V{q}} \\
 -\rmi\prefactor\rbr{\hat{a}_{\alpha}-\hat{a}_{\alpha}^\dagger}\ket{\V{p}}&=p_{\alpha}\ket{\V{p}},
\end{align}
with an arbitrary but fixed scale $\prefactor$.
These quadrature eigenstates obey the resolutions of unity
\begin{equation}
\intg{}{}{^Lq}\ket{\V{q}}\bra{\V{q}}=\hat{1}=\intg{}{}{^Lp}\ket{\V{p}}\bra{\V{p}}
\end{equation}
and their overlap matrix elements
are given by
\begin{align}
 \braket{\V{q}|\V{q}^\prime}&=\delta\rbr{\V{q}-\V{q}^\prime} \;, \\
 \braket{\V{p}|\V{p}^\prime}&=\delta\rbr{\V{p}-\V{p}^\prime} \;, \\
 \braket{\V{q}|\V{p}}&=\frac{1}{2\prefactor\sqrt{\pi}}\exp\rbr{\frac{\rmi\V{p}\cdot\V{q}}{2\prefactor^2}} \;.
\end{align}

Following the usual steps to derive the Feynman propagator 
(\textit{i.e.,} splitting up the exponential into a product of 
$N$ exponentials, inserting unity operators in terms of 
$\hat{\V{q}}$ and $\hat{\V{p}}$ between each pair of exponentials, 
and taking the limit $N\to\infty$), we find for the Hamiltonian
\begin{equation}
\hat{H}=\sum_{\alpha\beta}h_{\alpha\beta}\hat{a}_{\alpha}^{\dagger}
\hat{a}_{\beta}+\frac{1}{2}\sum_{\alpha\beta\mu\sigma}
V_{\alpha\beta\mu\sigma}\hat{a}_{\alpha}^{\dagger}\hat{a}_{\beta}
\hat{a}_{\mu}^{\dagger}\hat{a}_{\sigma}
\end{equation}
[see Eq.~(2) in the Letter] a path integral representation 
of the propagator in quadrature representation as
\begin{widetext}
\begin{eqnarray}
K\rbr{{\bf q}^{({\rm f})},{\bf q}^{({\rm i})},t}&=&
\lim\limits_{N\to\infty}\intg{}{}{{\bf q}^{(1)}}\cdots
\intg{}{}{{\bf q}^{(N-1)}}
\int\frac{{\rm d}{\bf p}^{(1)}}{4\pi\prefactor^2} \cdots
\int\frac{{\rm d}{\bf p}^{(N)}}{4\pi\prefactor^2}
\prodl{k=1}{N}\exp\Bigg[\frac{\rmi}{2\prefactor^2}
{\bf p}^{(k)}\cdot\rbr{{\bf q}^{(k)}-{\bf q}^{(k-1)}}
\nonumber \\ &&
-\frac{\rmi\tau}{\hbar}\sum_{\alpha,\beta=1}^L
h_{\alpha\beta} (\psi_\alpha^{(k)})^\ast \psi_\beta^{(k)}
-\frac{\rmi\tau}{2\hbar}
\sul{\alpha,\beta\mu,\sigma=1}{\sites} V_{\alpha\beta\mu\sigma}
(\psi_{\alpha}^{(k)})^\ast(\psi_{\mu}^{(k)})^\ast
\psi_{\beta}^{(k)}\psi_{\sigma}^{(k)}\Bigg],
\label{eq:descrete_propagator}
\end{eqnarray}
\end{widetext}
with 
$2\prefactor \gV{\psi}^{(k)}\equiv\V{q}^{(k-1)}+\rmi\V{p}^{(k)}$, 
$2\prefactor(\gV{\psi}^{(k)})^\ast \equiv \V{q}^{(k-1)}-\rmi\V{p}^{(k)
}$, and $\tau\equiv t/N$.
Calculating the integrals in stationary phase approximation and finally 
taking the limit $N\to\infty$ yields the van-Vleck-Gutzwiller 
propagator
\begin{eqnarray}
K\rbr{\qfv,\qiv,t}&=& \frac{1}{\rbr{-2\pi\rmi\hbar}^{\sites/2}}
\nonumber \\
&\times&
\sul{\gamma}{}\sqrt{\det\pdiff{^2R_\gamma}{\qiv\partial\qfv}}
\rme^{\rmi R_\gamma / \hbar}
\end{eqnarray}
where the sum runs over all possible classical trajectories 
defined by the equation of motion
\begin{eqnarray}
\rmi\hbar\pdiff{\psi_{\alpha}(s)}{s} &=& \sul{\beta=1}{\sites}
h_{\alpha\beta}\psi_{\beta}(s) \nonumber \\
&+& \sul{\beta,\mu,\sigma=1}{\sites}
V_{\alpha\beta\mu\sigma}\psi_{\mu}^\ast(s)
\psi_{\beta}(s)\psi_\sigma(s)
\end{eqnarray}
and the boundary conditions 
$2\prefactor {\rm Re}[\gV{\psi}(0)]=\qiv$ and
$2\prefactor {\rm Re}[\gV{\psi}(t)]=\qfv$.
$R_\gamma$ is the classical action given by
\begin{widetext}
\begin{equation}
R_\gamma\rbr{{\bf q}^{({\rm f})},{\bf q}^{({\rm i})},t}=
\intg{0}{t}{s}\Bigg\{\frac{\hbar}{2\prefactor^2}
{\bf p}(s)\cdot\dot{\bf q}(s)
-\sum_{\alpha,\beta=1}^L h_{\alpha\beta} \psi_\alpha^\ast(s)
\psi_\beta(s)
-\frac{1}{2}\sul{\alpha,\beta,\mu,\sigma=1}{\sites}
V_{\alpha\beta\mu\sigma} \psi_\alpha^\ast(s)
\psi_\beta(s)\psi_\mu^\ast(s)\psi_\sigma(s)\Bigg\}
\end{equation}
\end{widetext}
with $\V{q}(s)\equiv 2\prefactor{\rm Re}[\gV{\psi}(s)]$ 
and $\V{p}(s)\equiv 2\prefactor{\rm Im}[\gV{\psi}(s)]$ 
evaluated along the trajectory $\gamma$.

We now want to project the result onto Fock states. 
To this end, we need the overlap
\begin{equation}
\braket{n|q}=\frac{1}{\sqrt{2^nn!\sqrt{2\pi}\prefactor}}
\exp\rbr{-\frac{q^2}{4\prefactor^2}}
H_n\left(\frac{q}{\sqrt{2}\prefactor}\right)
\end{equation}
of a quadrature eigenstate $\ket{q}$ with a Fock state $\ket{n}$
associated with an individual single-particle orbital.
For large $n$, we can employ the WKB approximation for the Hermite polynomials $H_n$, 
which yields
\begin{eqnarray}
\braket{n|q}&\simeq&
\frac{\sqrt{2/\pi}}{(4\prefactor^2n-q^2)^{1/4}} \nonumber \\
&\times&\cos\cbr{\frac{q}{4\prefactor^2}
\sqrt{4\prefactor^2\rbr{n+1/2}-q^2}}
\end{eqnarray}
within the oscillatory region $\left|q\right|\leq 2\prefactor\sqrt{n}$.
Since $\braket{n|q}$ decreases exponentially for larger 
values of $|q|$, 
we can restrict the integration necessary for the projection onto 
Fock states to the oscillatory region $\left|q\right|\leq 2\prefactor\sqrt{n}$.

With this restriction, we substitute 
$q_{\alpha}^{({\rm i}/{\rm f})}\mapsto\theta_{\alpha}^{({\rm i}/{\rm f})}$ 
($\alpha\in\{1,\ldots,\sites\}$)
with $\thetaifj{\alpha}\in[-\pi,\pi]$ defined through
\begin{align}
q_1^{({\rm i})}&\equiv2\prefactor\sqrt{n_1^{({\rm i})}+1/2}
\cos\rbr{\theta^{({\rm i})}_1}, \\
q_1^{({\rm f})}&\equiv2\prefactor\sqrt{n_1^{({\rm f})}+1/2}
\cos\rbr{\theta_1^{({\rm f})}+\theta^{({\rm i})}_1},
\end{align}
as well as
\begin{align}
q_{\alpha}^{({\rm i}/{\rm f})}&\equiv2\prefactor\sqrt{n_{\alpha}^{({\rm i}/{\rm f})}+1/2}
\cos\rbr{\theta_{\alpha}^{({\rm i}/{\rm f})}+\theta^{({\rm i})}_1} \
\end{align}
for $\alpha=2,\ldots,\sites$.
The integrations over $\theta_1^{({\rm f})},\ldots,\theta_{\sites}^{({\rm f})},
\theta_2^{({\rm i})},\ldots,\theta_{\sites}^{({\rm i})}$ can now be performed in 
stationary phase approximation.
This selects trajectories that satisfy
\begin{align}
\psi_1(t)&=\sqrt{n_1^{({\rm f})}+1/2}
\exp\tbr{\rmi\rbr{\theta_1^{({\rm f})}+\theta^{({\rm i})}_1}}, \label{moduluscond1}
\end{align}
as well as
\begin{align}
\psi_{\alpha}(0)&=\sqrt{n_{\alpha}^{({\rm i})}+1/2}
\exp\tbr{\rmi\rbr{\theta_{\alpha}^{({\rm f})}+\theta^{({\rm i})}_1}} \label{moduluscond2}\\
\psi_{\alpha}(t)&=\sqrt{n_{\alpha}^{({\rm f})}+1/2}
\exp\tbr{\rmi\rbr{\theta_{\alpha}^{({\rm f})}+\theta^{({\rm i})}_1}} \label{moduluscond3}
\end{align}
for $\alpha=2,\ldots,\sites$.
Since the classical equations of motion preserve 
$|\psi_1(t)|^2 + \ldots + |\psi_L(t)|^2$, 
these stationary phase conditions already imply 
$|\psi_{1}(0)|^2=n_{1}^{({\rm i})}+1/2$ 
provided the final state $\ket{\textbf{n}^{({\rm f})}}$ contains as many 
particles as the initial state $\ket{\textbf{n}^{({\rm i})}}$.
Due to the $U(1)$ gauge symmetry, 
a variation of the global phase $\thetaij{1}$ 
will trivially give rise to another solution
$\boldsymbol\psi^\prime= \boldsymbol\psi \exp(\theta^{({\rm i})}_1)$ 
that exhibits the same initial and final populations as
$\boldsymbol\psi$. Therefore, one cannot solve the integral over $\thetaij{1}$ in stationary phase approximation, but needs to do it exactly, 
which yields 
an additional factor $2\pi$.

The propagator in Fock 
space then finally reads
\begin{eqnarray}
K\rbr{\nfv,\niv,t} &=& \frac{1}{\rbr{-2\pi\rmi\hbar}^{(\sites-1)/2}} \nonumber \\
& \times& \sul{\gamma}{}
\sqrt{{\det}^\prime\pdiff{^2R_\gamma}{\niv\partial\nfv}}
\rme^{\rmi R_\gamma / \hbar},
\end{eqnarray}
where $\gamma$ indexes all classical trajectories that satisfy the boundary conditions
$\left|\psi_{\alpha}(0)\right|^2 = n_{\alpha}^{({\rm i})}+1/2$
and $\left|\psi_{\alpha}(t)\right|^2 = n_{\alpha}^{({\rm f})}+1/2$
for $\alpha=1,\ldots,L$
with the global phase $\theta_1^{({\rm i})}$ being fixed through the choice
$\psi_1(0)=[\nij{1}+1/2]^{1/2}$.
The prime at the determinant 
\begin{equation}
{\det}^\prime\pdiff{^2R_\gamma}{\niv\partial\nfv} \equiv
\det\left(\pdiff{^2R_\gamma}{\nij{\alpha}\partial\nfj{\beta}}
\right)_{\alpha,\beta = 2,\ldots,L}
\end{equation}
indicates that it is taken with respect to the matrix obtained by skipping the derivatives with respect to the first components.

\end{document}